\documentclass[reprint,showpacs,showkeywords,aps,prb,twocolumn,unsortedaddress,superscriptaddress]{revtex4-1}

\usepackage[T1]{fontenc}
\usepackage{graphicx}
\usepackage{dcolumn}
\usepackage{bm}
\usepackage{hyperref}
\usepackage{color}
\usepackage{epsfig}
\usepackage{bm}

\begin{document}

\title{Chiral properties of hematite  ($\alpha$$-$Fe$_2$O$_3$) inferred from resonant Bragg diffraction using circularly polarized x-rays }

\author{A. Rodr\'{\i}guez-Fern\'andez }
\author{J. A. Blanco }
\affiliation{Departamento de F\'{\i}sica, Universidad de Oviedo, E-33007 Oviedo, Spain }

\author{S. W. Lovesey}
\affiliation{ ISIS Facility $\&$ Diamond Light Source Ltd,
Oxfordshire OX11 0QX, United Kingdom}

\author{V. Scagnoli }
\author{U. Staub }
\affiliation{Paul Scherrer Institut, 5232 Villigen, Switzerland }

\author{H. C. Walker}
\author{D. K. Shukla }
\author{J. Strempfer }
\affiliation{Deutsches Elektronen-Synchrotron, 22607 Hamburg, Germany }

\date{\today}

\begin{abstract}
Chiral properties of the two phases -- collinear motif (below Morin transition temperature, T$_M$ $ \approx \rm 250$ $K$) and canted motif (above T$_M$) -- of magnetically ordered hematite ($\alpha$$-$Fe$_2$O$_3$) have been identified in single crystal resonant x-ray Bragg diffraction, using circular polarized incident x-rays tuned near the iron K-edge. Magneto-electric multipoles, including an anapole, fully characterize the high-temperature canted phase, whereas the low-temperature collinear phase supports both parity-odd and parity-even multipoles that are time-odd. Orbital angular momentum accompanies the collinear motif, while it is conspicuously absent with the canted motif. Intensities have been successfully confronted with analytic expressions derived from an atomic model fully compliant with chemical and magnetic structures. Values of Fe atomic multipoles previously derived from independent experimental data, are shown to be completely trustworthy. 
\end{abstract}


\maketitle

\section{Introduction}
\begin{figure}[h]
\begin{center}
\includegraphics[width=8cm, angle=270]{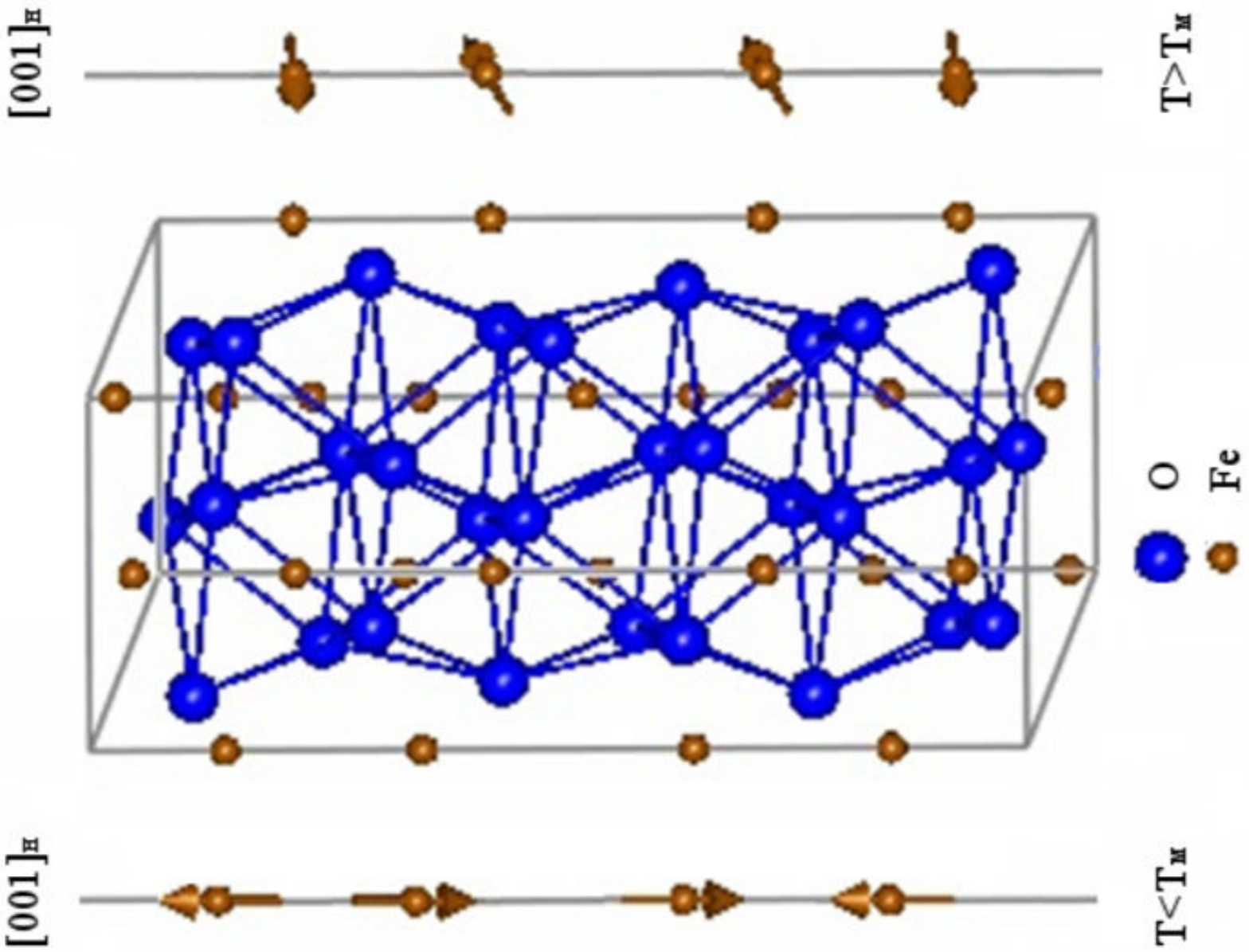}
\caption{\label{fig:figure1}(Color Online) Crystal and magnetic structure of hematite. The orange small dots represent ferric ions while the blue large dots present the oxygen atoms positions. The up line denotes the magnetic motif along the $c$-axis below the Morin temperature (phase I). The down line denotes the motif above the Morin temperature, where iron moments are contained in the $a$-$b$ plane (phase II).}
\end{center}
\end{figure}

Alpha-ferric oxide ($\alpha$$-$Fe$_2$O$_3$), also known as hematite, a name deriving from the Greek  \"{ }$\alpha$$\iota$$\mu$$\alpha$$\tau$$\iota$$\tau$$\eta$$\varsigma$\"{ }  due to its blood-like shade in powder form, is still today revealing its mysteries. \cite{Morrish1, Catti2} hematite has been present in scientific literature since the studies performed by the Greek philosopher Theophrastus around 315 B.C and, later was studied by the father of magnetism, William Gilbert of Colchester in the 16th century. Its magnetic behavior was first studied in the early twentieth century by Honda and Son\'{e} (1914), but it was not until Dzyaloshinsky in 1958 when it was defined as a canted antiferromagnet, becoming the prototype of the Dzyaloshinsky-Moriya interaction. \cite{Dzyaloshinsky3, Moriya4}

hematite is a member of the corundum-structure family (centro-symmetric space-group $\sharp 167$, $R\bar{3}c$) and the ferric ($Fe^{3+}$, $3d^5$) ions present in $\alpha$--Fe$_2$O$_3$ are arranged along the $c$-axis occupying 4(c) sites. The antiferromagnetic behavior present in this compound below its N\'{e}el Temperature ( T$_N$ $ \approx \rm 948$ $K$) shows two different magnetic orders separated by the Morin transition temperature,  T$_M$ $ \approx \rm 250$ $K$. Below this temperature the magnetic moments are all parallel to the hexagonal $c$-axis in a collinear antiferromagnetic G-type configuration (that is, the nearest neighbors have opposite spins while the next-nearest neighbors have parallel spins) with an iron magnetic moment $= 4.9$ $\mu_B$ at $77$ $K$, while above T$_M$ the material shows a magnetic motif where the moments are in a (basal) plane normal to the $c$-axis showing a canted antiferromagnetic order, depicted in Figure \ref{fig:figure1}. As previously, we follow Dzyaloshinsky and label the collinear (low-temperature phase) and canted (room-temperature phase) magnetic motifs as I and II, respectively.\cite{Lovesey5} The Dzyaloshinsky-Moriya antisymmetric interaction is responsible for the behavior known as weak ferromagnetism that in the case of the hematite is parallel to the diad axis of rotation symmetry.\cite{Morrish1, Dzyaloshinsky3}

Chiral order in electronic structure is unambiguously detected by a probe with a matching characteristic, as discussed in Rodr\'{i}guez-Fern\'{a}ndez$ \it et \, al. $,\cite{Rodriguez6} and we have used circularly polarized x-rays to verify the existence of such order in a single crystal of hematite. No previous work of this type has been published for hematite, to the best of our knowledge. Tuning the x-ray energy to an atomic resonance of a ferric ion, the Fe K-edge, means these ions and no others participate in the chiral order observed. In addition, the resonant process enhances the sensitivity of the scattering process to the local environments and angular anisotropy in the electron distribution that appears due to the spin, charge or multipolar order.\cite{Lovesey5, Rodriguez6, Lovesey7, Fernandez8,Lovesey9,Wilkins10,Paixao11,Tanaka12}

Previous experiments on hematite, using Bragg diffraction of linearly polarized x-rays with the primary energy tuned near the iron K-absorption edge, were performed by Finkelstein $ \it et \, al. $  and Kokubun $ \it et \, al. $.\cite{Finkelstein13, Kokubun14} Due to the important contribution of the Thomson scattering in these types of experiments, attention is given to Bragg reflections that are space-group forbidden by extinction rules. Integer Miller indices obey the extinction rule $ l$ odd and $-h + k + l = 3n$ in the case of hematite. 

Finkelstein $ \it et \, al. $  observed,\cite{Finkelstein13} while rotating a single crystal about the Bragg wavevector $(003)_H$ in a so-called azimuthal-angle scan,  a six-fold periodicity of the intensity that is traced to a triad axis of rotation symmetry that passes through sites occupied by resonant, ferric ions. At a later date, Kokubun $ \it et \, al. $ reported azimuthal-angle scans for $l = 3$ and $9$.\cite{Kokubun14} Unlike these two groups, we exploit polarization analysis of the diffracted beam to unveil contributions to the magnetization with different spatial symmetries. On the way, we confirm our prediction that hematite supports chiral order,\cite{Lovesey5} and gain confidence in our previously reported values of Fe atomic multipoles, because they provide a totally satisfactory description of new azimuthal-angle data for $l = 3$ and $9$ gathered in phase I and phase II.  A potential uncertainty in our analysis Ref. (\onlinecite{Lovesey5}) of data reported in Ref. (\onlinecite{Kokubun14}) is set to rest. Previously, we were forced to the conclusion that there is an error in Ref. (\onlinecite{Kokubun14}) on the reported setting of the crystal in azimuthal-angle scans, and the error is confirmed here by use of our own data.

In this communication, we present data from a circular polarized x-ray diffraction experiment performed at the Fe K-edge. Section 2 contains the description of the crystal and experiment. This is followed by the discussion of the results in the Section 3, where we report a detailed analysis of our azimuthal angle scans for $l = 3$ and $9$, for the hematite sample at $150\ K$ (collinear motif, phase I) and $300 \ K$ (canted motif, phase II). In Section 4 we present our final remarks and conclusions.

\section {Crystal and experimental method}

The synthetic hematite single crystal studied in this experiment was purchased from the Mateck Company. The size of the sample was about 10 $\times$ 10 mm$^2$ with a thickness of 0.5 mm, showing a polished surface near the $[00l]_H$ direction. In Cartesian coordinates our hexagonal crystal coordinates are a$_H$ $=$ a$(1, 0, 0)$, b$_H$ $=$ a$(-1/2, \surd 3/2, 0)$ and c$_H$ $=$ c$(0, 0, 1)$, with a $= 5.038$ \AA  \ and c $= 13.712$ \AA. 

\begin{figure}[h]
\begin{center}
\includegraphics[width=8cm]{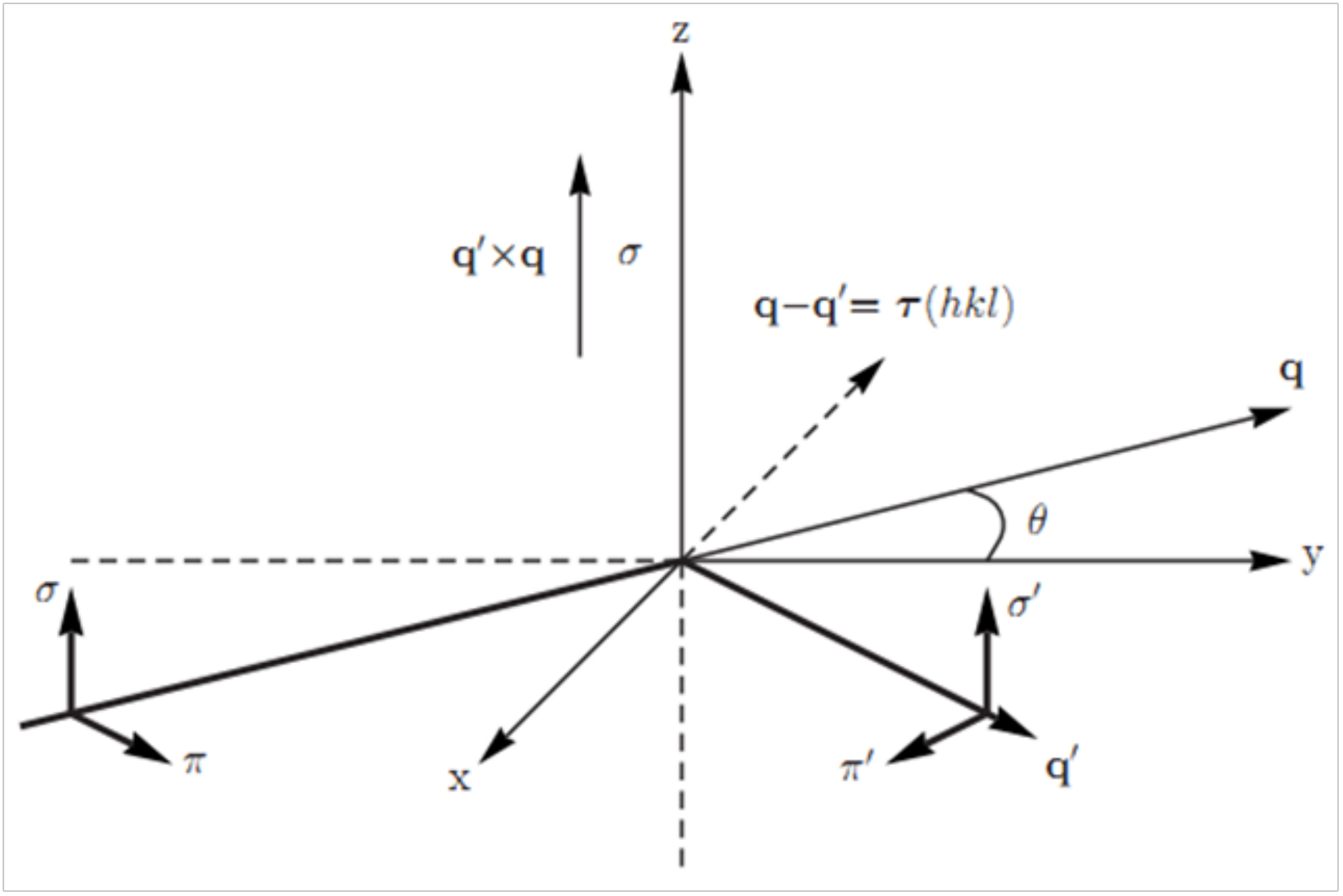}
\caption{\label{fig:figure2}Cartesian coordinates (x, y, z) and x-ray polarization and wavevectors. The plane of scattering spanned by primary ($\bf q$) and secondary ($\bf q'$) wavevectors coincides with the x-y plane. Polarization labeled $\sigma$ and $\sigma'$ is normal to the plane and parallel to the $z$-$axis$, while polarization labeled $ \pi$ and $\pi'$ lies in the plane of scattering. The beam is deflected through an angle $2\theta$.}
\end{center}
\end{figure}

The experimental data presented in this work were obtained at the Beamline P09, located in the synchrotron source PETRA III (Germany). \cite{Strempfer15} This beamline covers the energy range from $2.7$ to $24$  $keV$. A double phase$-$retarder setup is used to obtain the circular and rotated linear polarization for the incident beam. The double phase-retarder setup corrects for some depolarizing effects and accomplishes a better rotated polarization rate. \cite{Francoual16,Okitsu17,Scagnoli18}  Details of incident polarization manipulation using diamond phase plates at P09 are described elsewhere. \cite{Strempfer15,Francoual16} The phase plates are followed by a focusing and higher harmonic rejection system consisting of vertically reflecting mirrors. The plate shaped crystal, attached to the cold finger of a closed-cycle cryostat, was mounted on a Psi-diffractometer such that the $[0 0 l]_H$ direction of the crystal is parallel to the scattering vector, ($\bf q- q$'), as shown in Figure \ref{fig:figure2}. Polarization analysis has been performed using Cu$(2 2 0)$ analyzer crystal. States of polarization labeled $\pi$ ($\pi$') and $\sigma$ ($\sigma$') are defined in Figure \ref{fig:figure2}. 

\begin{figure}[h]
\begin{center}
\includegraphics[width=8cm]{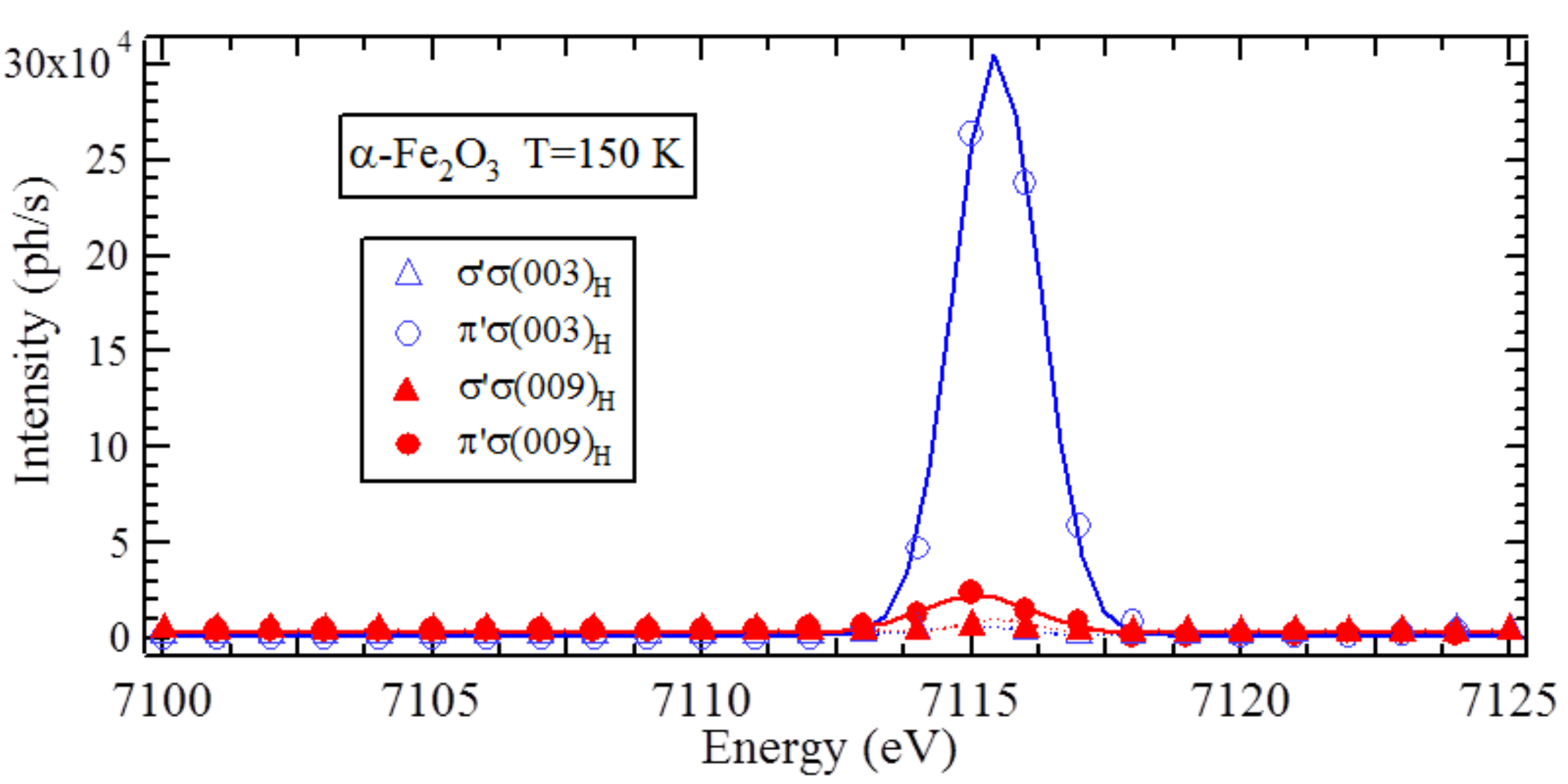}
\caption{\label{fig:figure3}(Color Online) X-ray energy dependence below T$_M$ $(T=150 K)$ for both reflections $(003)_H$ and $(009)_H$ in the vicinity of the iron K-edge. The blue empty dots (triangles) show the linear polarized $\pi$'$\sigma$ ($\sigma$'$\sigma$) data for the the $(003)_H$, while the red dots (triangles) define the data from the $\pi$'$\sigma$ ($\sigma$'$\sigma$) polarized channel for the $(009)_H$. The solid lines present the fitting to a model of a single oscillator.}
\end{center}
\end{figure}

In the case of $R\bar{3}c$, the reflections $(003)_H$ and $(009)_H$ are space-group forbidden, but weak Bragg diffraction occurs near an atomic resonance, as demonstrated by data displayed in Figure \ref{fig:figure3}. In the experiment performed at beamline P09, the energy at which the primary x-ray beam was tuned, $7115$ $eV$, is close to the iron K-edge. At this energy the forbidden $(003)_H$ and $(009)_H$ reflection were investigated with the sample maintained at two different temperatures, below ($150$ $K$) and above ($300$ $K$) the Morin temperature.

During the experiment the incident polarization was switched between almost perfect right and left circular polarization. A measure of the high quality of circular polarization of the primary beam is demonstrated by small values of the parameters for linear polarization (following the convention of Pauli matrices as done by Lovesey $ \it et \, al. $), namely, P$_1$ $= 0.010 \ \pm \ 0.002$ and P$_3$ $= 0.026 \ \pm \ 0.002$ for right-handed, and P$_1$ $= -0.016 \ \pm \ 0.002$ and P$_3$ $= 0.036 \ \pm \ 0.002$ for left-handed (properties of Stokes parameters are mentioned again in Section 3). \cite{Fernandez19, Lovesey20}

We have found an extensive contribution from Renninger reflections, also known as multi-beam reflections. The subtraction of this kind of back-ground intensity was done using a Matlab program developed by Gareth Nisbet as previously done for the extraction of the data present in Ref. (\onlinecite{Rodriguez6}). Attention was focused on azimuthal angles either only lightly, or not contaminated by Renninger reflections (therefore measured points in the azimuth dependence are not equidistant). 

\begin{figure}[h]
\begin{center}
\includegraphics[width=8cm]{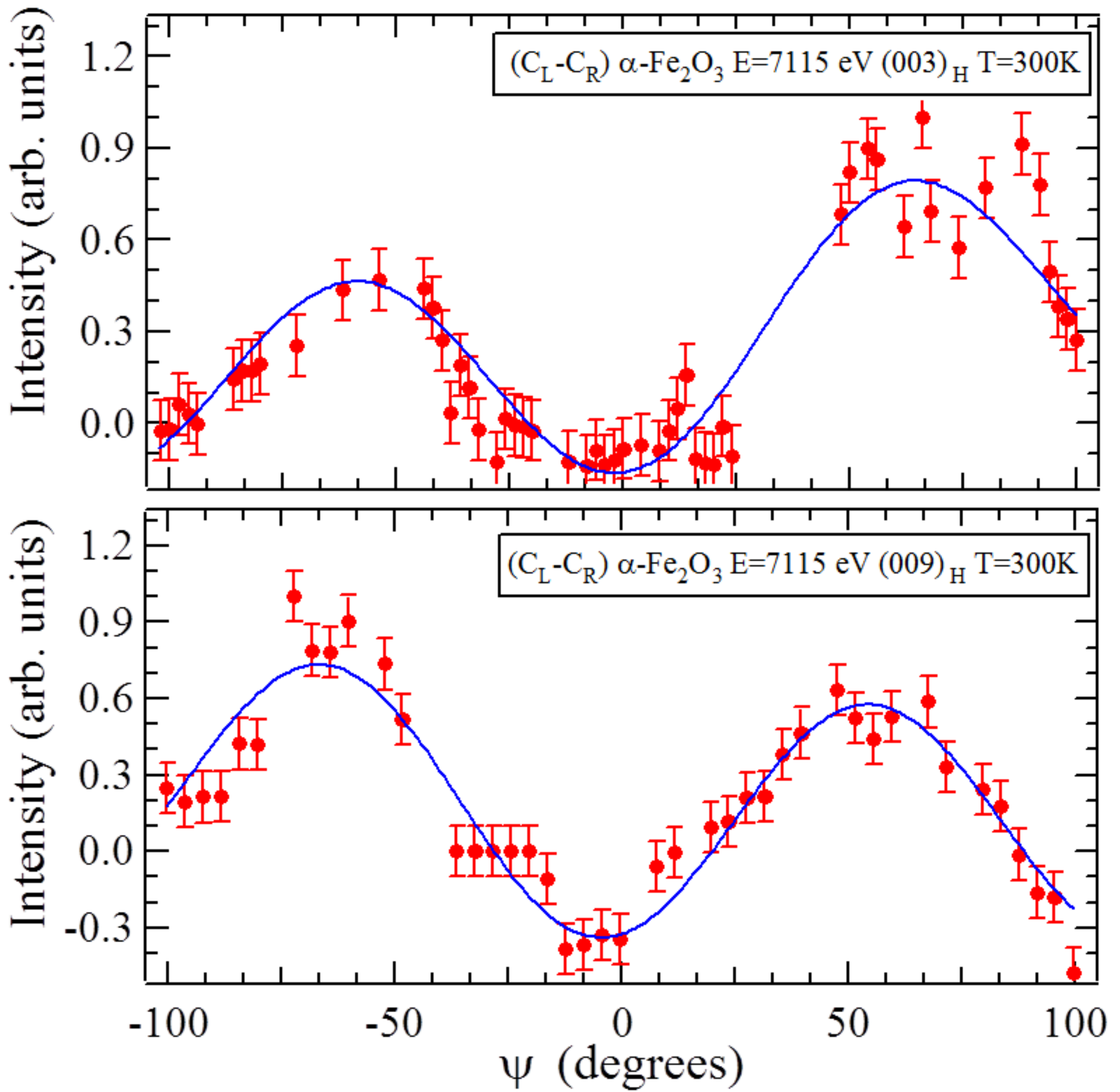}
\caption{\label{fig:figure4}(Color Online) Azimuthal-angle scans for phase II (canted motif) at $300$ $K$. Difference between circular left (C$_L$) and circular right (C$_R$) polarization for the $(003)_H$ (upper panel) and the $(009)_H$ (lower panel). The red dots represent the experimental data while the blue line shows expression (\ref{Eq5}) for pure E1$-$E2 resonance evaluated with multipoles taken from Ref. (\onlinecite{Lovesey5}) and reproduced in Table I.}
\end{center}
\end{figure}

Circular left (C$_L$) and circular right (C$_R$) polarized azimuthal scans were performed at room temperature; the difference between these two polarizations is presented in Figure \ref{fig:figure4}. Fitting to data above T$_M$ was performed using equations (\ref{Eq3}) and (\ref{Eq4}) presented in section 3. The multipole values used for these fittings are shown in Table I (phase II) and they agree with the ones derived by Lovesey $ \it et \, al. $ \cite{Lovesey5}

\begin{figure}[h]
\begin{center}
\includegraphics[width=8cm]{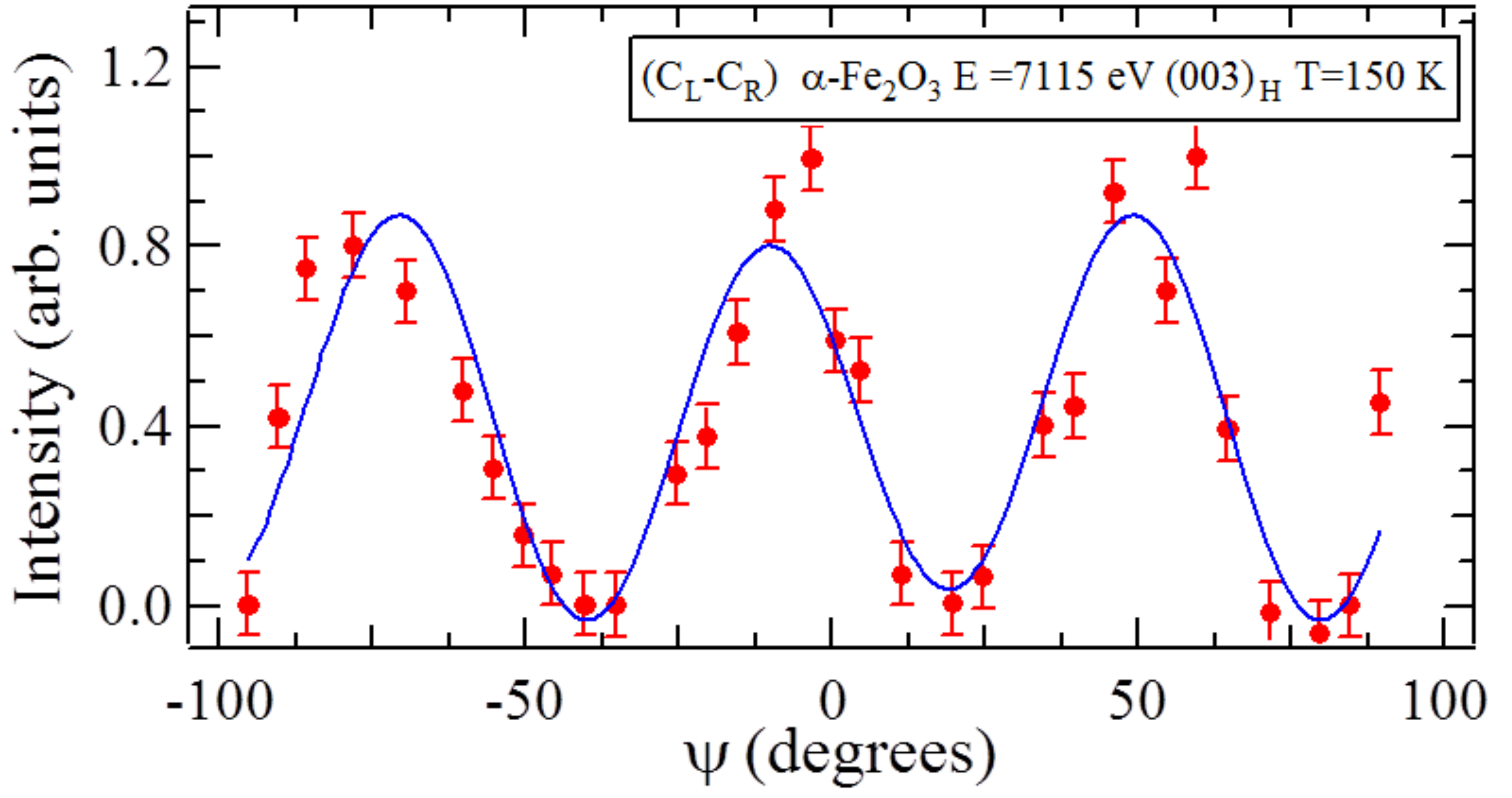}
\caption{\label{fig:figure5}(Color Online) Azimuthal-angle scan for phase I (collinear motif) at $150$ $K$. Difference between circular left (C$_L$) and circular right (C$_R$) polarization for $(003)_H$. The red dots represent the experimental data while the blue line shows expression (\ref{Eq3}) and (\ref{Eq4}) for a mixture of E1$-$E2 and E2$-$E2 evaluated with multipoles taken from Ref. (\onlinecite{Lovesey5}) and reproduced in Table I. In (\ref{Eq4}) the parity-even and time-odd octupole $\langle T^3_3 \rangle"$ is given a nominal value $1 (\pm0.1) 10^{-4}$. }
\end{center}
\end{figure}

\begin{figure}[h]
\begin{center}
\includegraphics[width=8cm]{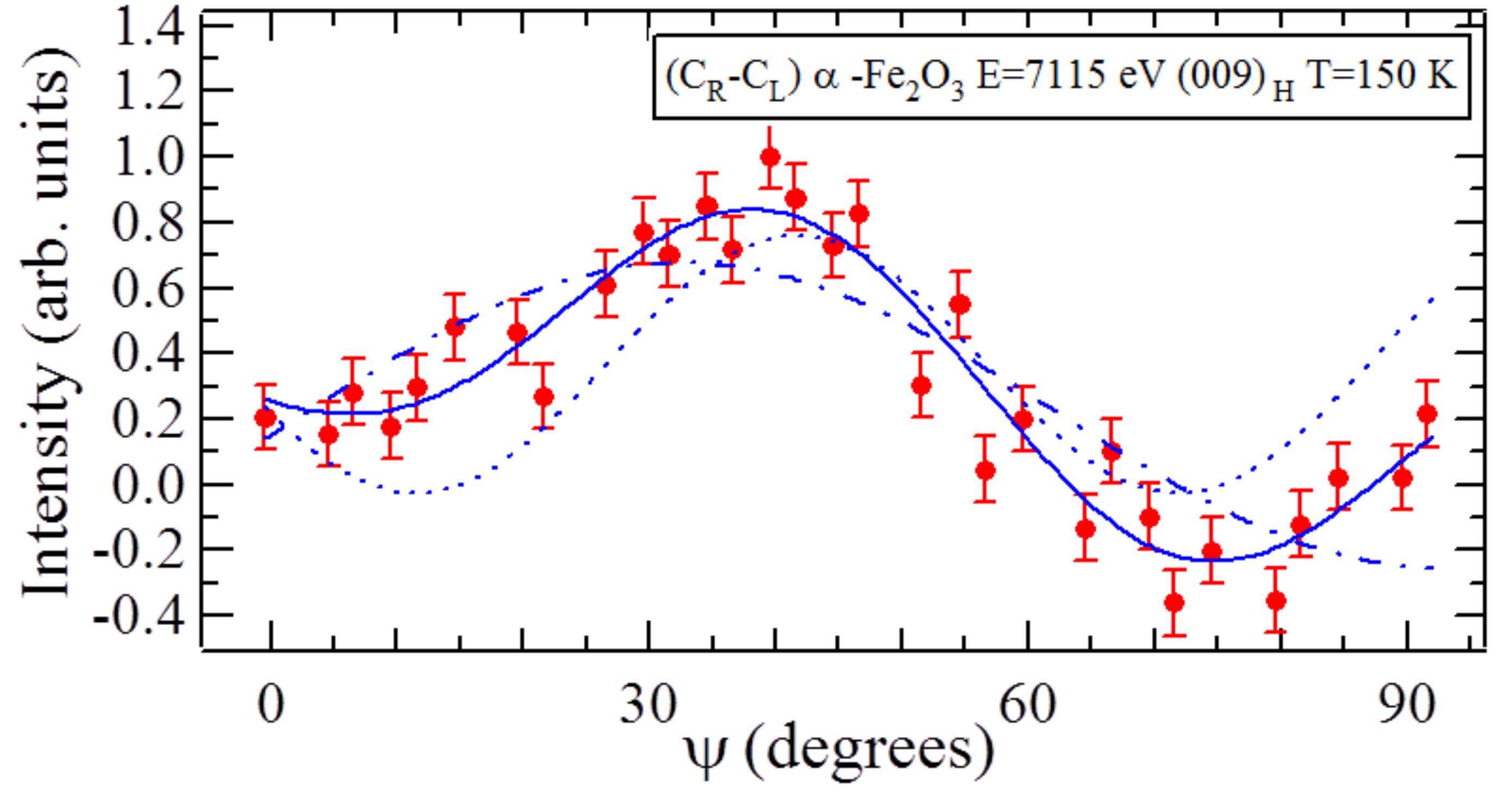}
\caption{\label{fig:figure6}(Color Online) Azimuthal-angle scan for phase I (collinear motif) at $150$ $K$. Difference between circular left (C$_L$) and circular right (C$_R$) polarization for the $(009)_H$.As in Figure \ref{fig:figure5} the red dots represent the experimental data while the blue line shows a mixture of E1-E2 and E2-E2. Data are compared to pure E1-E2 (\ref{Eq3}) (dashed line) and pure E2-E2 (\ref{Eq4}) (dotted line), with multipoles taken from Ref. (\onlinecite{Lovesey5}) and reproduced in Table I. In (\ref{Eq4}) the octupole $\langle T^3_3 \rangle"$ is given a nominal value 0.0001. }
\end{center}
\end{figure}

The azimuthal scan dependence of the $(003)_H$ reflection below T$_M$ is presented in Figure \ref{fig:figure5}.  While for  $(009)_H$ reflection is shown in  Figure \ref{fig:figure6}. As in the case of room temperature, multipole values used in the fitting are those derived by Lovesey $ \it et \, al. $, collected in Table I (phase I).\cite{Lovesey5} 

\section {Results and discussion}

For an interpretation of the experimental data, shown in Figures \ref{fig:figure4},  \ref{fig:figure5} and  \ref{fig:figure6} , we proceed as in Ref. (\onlinecite{Lovesey5}). The contribution of Thomson scattering is absent at space-group forbidden reflections, leaving a sum of non-resonant spin and a resonant contribution as ingredients for the appropriate scattering amplitude. \cite{Lovesey20,Lovesey21, Urs22, Urs23} 

The spin contribution, G$^s$, is explicitly first order in the small quantity $(E/mc^2)$, where $E$ is the primary energy and $mc^2$ the electron rest mass energy. Using notation displayed in Figure \ref{fig:figure2}, $G^s = i(E/mc^2)(\bf e \times \bf e') \bullet \bf {F_s(k})$, where $\bf {k = q - q'} $ $= (h, k, l)$ and $\bf e$ and $\bf {q (e'}$ and $\bf q')$ are, respectively, the polarization vector and wavevector of the primary (secondary) photon, while $\bf {F_s(k)}$ is the unit-cell structure factor for spin dipoles.

The measured energy profiles, displayed in Figure \ref{fig:figure3} for the reflections $(003)_H$ and $(009)_H$, show a single resonance in the pre-edge region that can be adequately modeled by a single oscillator centered at an energy $\Delta = 7115$ $eV$.\cite{Lovesey5} The resonant contribution to scattering is represented by $d(E) F_{\mu'\nu}$, where $d(E) = \Delta/[E - \Delta+ i\Gamma]$ with  $\Gamma$ the width in energy and $F_{\mu'\nu}$  a unit-cell structure factor for states of polarization $\mu$' (secondary) and $\nu$ (primary), as in Figure \ref{fig:figure2}. 

The generic form of our Bragg scattering amplitude for hematite at a space-group forbidden reflection (no Thomson scattering) is,
\begin{equation}
 G_{\mu'\nu}  (E) = G^s_{\mu'\nu}+ \rho \, d(E) \, F_{\mu'\nu},
\label{Eq1}
\end{equation}

In this expression, $\rho$ is a collection of factors, which include radial integrals for particular resonance events, namely, $\rho(E1-E1) = [\{R\}_{sp}/a_o]^2\aleph$, $\rho(E1-E2) = [q\{R^2\}_{sd}\{R\}_{sp}/a_o^2]\aleph \ and \ \rho(E2-E2) = [q\{R^2\}_{sd}/a_o]^2\aleph$. Here, $\aleph$ is a dimensionless quantity related to the Bohr radius $a_0$ and the resonant energy $\Delta = 7115$ $eV$, with $\aleph = m \Delta a_o^2/\hbar^2 = 260.93$. The sizes of radial integrals for the E1 and E2 processes at the K-absorption edge, $\{R\}_{sp}$ and $\{R^2\}_{sd}$, are discussed in Ref. (\onlinecite{Lovesey5}).

The polarization state of the photons, already briefly discussed in Section 2, is defined by Stokes parameters that are purely real and time-even, namely, ordinary scalars P$_1$ and P$_3$ for linear polarization, and a pseudo-scalar P$_2$ that represents the helicity of the beam. The contribution to the total intensity induced by circular polarization (helicity), I$_c$, is, \cite{Fernandez19}
\begin{equation}
 I_c = P_2 \,Im \,\{G^*_{\sigma'\pi}\, G_{\sigma'\sigma} + G^*_{\pi'\pi}\, G_{\pi'\sigma}\},
\label{Eq2}
\end{equation}

where the amplitudes G$_{\mu '\nu}$ are given by (\ref{Eq1}) and * denotes complex conjugation.  I$_c$ is zero for Thomson scattering since it is proportional to   $(\bf e . \bf e')$ and diagonal with respect to the polarization states.

We make use of unit-cell structure factors reported in our previous publication.\cite{Lovesey5}  Full use is made of the established chemical and magnetic structures in their construction. Degrees of freedom in the electronic ground-state of a ferric ion are captured in atomic multipoles labeled by their rank, $K$, and projection $Q ( -K \leq Q \leq K)$. \cite{Lovesey20,Lovesey21} Two types of multipoles are required, parity-even, $\langle T^K_Q \rangle$, and two flavors of parity-odd multipoles, $\langle G^K_Q \rangle$ and $\langle U^K_Q \rangle$, distinguished by their time-signatures. Magneto-electric multipoles, $\langle G^K_Q \rangle$, are time-odd and absent in the paramagnetic phase, and polar multipoles, $\langle U^K_Q \rangle$, are time-even, while $(- 1)^K$  is the time signature of $\langle T^K_Q \rangle$. Parity-even multipoles arise in E1-E1 and E2-E2 resonant events, and parity-odd multipoles are required for E1-E2, where E1 denotes an electric-dipole operator and E2 denotes an electric-quadrupole operator. All multipoles have the complex conjugate $\langle O^K_Q \rangle$* $=   (- 1)^Q$  $\langle O^K_{-Q} \rangle$, with $\langle O^K_0 \rangle$ purely real, and the relative phase of real and imaginary components is set by $\langle O^K_Q \rangle$ = $\langle O^K_Q \rangle$' $+ i\langle O^K_Q \rangle$".

Expressions for I$_c$ given in Ref. (\onlinecite{Lovesey5}) are repeated here for the convenience of the reader. 
(A) Below T$_M$ (phase I, collinear motif). There is no contribution to I$_c$ from E1$-$E1, due to crystal symmetry, and,

\begin{eqnarray}
 I_c(E1-E2)= -P_2\,(\frac {8 \sqrt{2}}{5})\rho^2 (E1-E2) \mid d(E)\mid ^2 sin(3 \psi)\,\times \nonumber\\
 \times cos^3 (\theta)\,(1+sin^2(\theta))cos^2 (\varphi l) \langle G^3_{+3}\rangle'\langle U^2_0 \rangle,\nonumber\\
\label{Eq3}
\end{eqnarray}

\begin{eqnarray}
 I_c(E2-E2)= -P_2 \,4 \rho ^2 (E2-E2) \mid d(E) \mid ^2 \,sin(6 \psi) \times \nonumber\\
 \times sin(\theta)\,cos^6 (\theta)\,sin^2 (\varphi l)\langle T^3_{+3}\rangle''  \langle T^4_{+3} \rangle',\nonumber\\
\label{Eq4}
\end{eqnarray}

In these expressions, the angle $\varphi$ $= -\pi u$, where $u=2z-1/2=0.2104$ for $\alpha$$-$Fe$_2$O$_3$. 

At the K-edge, parity-even multipoles with K odd are functions only of orbital angular momentum, in the electronic ground-state of the resonant ion-spin degrees of freedom are absent. \cite{Lovesey24} In which case, $\langle T^3_3 \rangle$" in (\ref{Eq4}) is zero for a pure ferric ion, because it has a shell that is half filled and spherically symmetric ($^6$S, $3d^5$).  In our previous study, where we interpreted data published by Kokubun $ \it et \, al. $, \cite{Kokubun14} we took $\langle T^3_3 \rangle$" $= 0$ on this basis. Our superior data displayed in Figures  \ref{fig:figure5} and  \ref{fig:figure6}, collected with the benefit of polarization analysis, shows that $\langle T^3_3 \rangle$" is different from zero in (\ref{Eq4}). In consequence, the ferric ion possesses unquenched orbital angular momentum. 

Dashed and dotted lines in Figure \ref{fig:figure6} shows our data for the (009)$_H$  reflection compared separately to E1-E2 and E2-E2. Evidently, a single event is not responsible for our observed intensities. However, a combination of the two events, E1-E2 and E2-E2 , provides a satisfactory account, with the fit represented by the continuous line in  Figure \ref{fig:figure6} confirms that this is so.

Concerning the contribution to the intensity from parity-odd multipoles (\ref{Eq3}), the requirement to have a value different from zero tells us that ($\langle G^3_3 \rangle$', $\langle U^2_0 \rangle$) is not zero. Notably, $\langle U^2_0 \rangle$ is a manifestation of local chirality.\cite{Lovesey20, Lovesey21, Lovesey24, Dmitrienk25}. The polar quadrupole is the same in the two phases, because it is related to chemical structure, while $\langle G^3_3 \rangle$' has a similar, small value in phase I, and a much larger value in phase II.

(B) Above T$_M$ (phase II, canted motif). Our data for this phase and two reflections are displayed in Figure \ref{fig:figure4}. Appropriate expressions for intensities induced by circular polarization in the primary beam are,\cite{Lovesey5}

\begin{widetext}
\begin{eqnarray}
I_c(E1-E2)= P_2 \,(\frac {8 \sqrt{2}}{5})\,\rho^2 (E1-E2)\,\mid
d(E)\mid ^2 \,cos^2(\varphi l)\, cos^2(\theta ) \,\langle U^2_0\rangle \times \nonumber\\
 \times  \{\frac{1}{\sqrt{3}} \,sin(\psi)\,[\frac{-3}{\sqrt{5}}\,(cos(3 \theta)+cos(\theta))\,\langle
G^1_{+1} \rangle' (cos(3 \theta)-cos(\theta)) \, \langle
G^2_{+1}\rangle''-\nonumber\\ - \frac{1}{\sqrt{5}}\,
(cos^3(\theta)+2cos(\theta))\, \langle G^3_{+1}\rangle']
-sin(3\psi)\,cos(\theta)\,(1+sin^2(\theta)) \langle G^3_{+3}\rangle'\},\nonumber\\
 \label{Eq5}
\end{eqnarray}

\begin{eqnarray}
I_c(E2-E2)=-P_2(\frac {1} {\sqrt{2}}) \,  \rho^2 (E2-E2)\, \mid
d(E)\mid ^2 \,sin^2(\varphi l)\, \langle T^4_{+3}\rangle'\times \nonumber\\
 \times \{ 4\,sin(\psi)\,cos^4(\theta)\,[\frac{-1}{\sqrt{5}}\,
sin(\theta)\,(8cos^2(\theta)-5) \langle T_1^1 \rangle''
+\nonumber\\ +\sqrt{\frac{3}{5}}\, sin(\theta )cos^3 (\theta)\langle
T^3_{+1}\rangle'']-4\sqrt{2}\,sin(\theta)\,cos^6(\theta)\,sin(6\psi)\,\langle
T^3_{+3} \rangle''\},\nonumber\\ 
\label{Eq6}
\end{eqnarray}
\end{widetext}

Our data in Figure \ref{fig:figure4} agree with the prediction of an E1-E2 event (\ref{Eq5}) evaluated with multipoles carried over from our previous work. \cite{Lovesey5} Correspondingly, magneto-electric multipoles are large compared to their values in phase I, with an octupole dominant. Treating $\langle T^K_Q \rangle$ with $K$ odd in (\ref{Eq6}) as unknowns, it is not possible to find a satisfactory fit to a pure E2-E2 event, and it has no role in an interpretation of phase II. As the hexadecapole $\langle T^4_3 \rangle$' is the same in the two phases, because it is determined by chemical structure, likewise local chirality $\langle U^2_0 \rangle$, we conclude that orbital angular momentum, manifest through $\langle T^K_Q \rangle$ with K odd, is insignificant in the high-temperature magnetic phase.

\begin{table}[h]
\caption{\label{tab:table1}Numerical values of multipoles reported in Ref. (\onlinecite{Lovesey5}) and used here for intensities generated from expressions (\ref{Eq3}) - (\ref{Eq6}). As in Ref. (\onlinecite{Lovesey5}), $\langle T^4_3 \rangle$' and $\langle U^2_0 \rangle$, multipoles, which contribute in both phases, are fixed to $10$ and $0.5$, respectively.}
\begin{ruledtabular}
\begin{tabular}{@{}lcc}

    Multipole                       &     Phase I    &   Phase II      \\
\colrule
 $\langle G^1_{+1} \rangle '  $     &     $           -                   $   &       $5.0 (2) 10^{-1} $    \\
 $\langle G^2_{+1} \rangle '' $     &     $           -                   $   &       $-3.8 (3)10^{-1} $    \\
 $\langle G^3_{+1} \rangle '  $     &     $           -                   $   &      $ 10.7 (6)10^{-1} $    \\
 $\langle G^3_{+3} \rangle '  $     &     $  4.1 (2) 10^{-1}    $   &      $24.5 (5)10^{-1} $    \\
 $\langle T^3_3 \rangle "        $     &     $ 1.0 (1) 10^{-4}     $   &      $       -                       $    \\

\end{tabular}
\end{ruledtabular}
\end{table}

\section {Conclusion}

We report extensive data on magnetically ordered hematite gathered with the experimental technique of x-ray Bragg diffraction augmented by an atomic resonance. The primary energy was tuned close to the iron K-edge, and intensities measured at space-group forbidden reflections, $(003)_H$ and $(009)_H$, that are exceptionally sensitive to magnetic degrees of freedom in the electronic ground-state. Use of polarization analysis improved the quality of our data significantly. We chose circular polarization, and report differences in intensities gathered with left and right-handed primary polarization. 

The existence of intensity induced by circular polarization confirms that magnetically ordered hematite is chiral, as we predicted. \cite{Lovesey5} Moreover, we confirm that our previous estimates of parity-odd multipoles, using data published by Kokubun $ \it et \, al. $, \cite{Kokubun14} are completely trustworthy. Below the Morin transition, the collinear motif contains orbital angular momentum and the ferric ion is not spherically symmetric, e.g., $^6$S, $3d^5$. However, we find no evidence of orbital angular momentum in the canted motif that exists above the Morin transition. In this phase, diffraction can be interpreted with parity-odd multipoles only, with magneto-electric octupoles making the dominant contribution.

\section{ACKNOWLEDGMENTS}
Financial support has been received from Spanish FEDER-MiCiNN Grant No. MAT2011-27573-C04-02. One of us A.R.F is grateful to Gobierno del Principado de Asturias for the financial support from Plan de Ciencia, Tecnolog\'ia e innovaci\'on (PTCI) de Asturias.  We thank the Deutsches Elektronen-Synchrotron for access to beamline P09 of the synchrotron PETRA III (I-20110433 EC) that contributed to the results presented here. This work has been supported by National Centre of Competence of Research-Materials with Novel Electrical Properties of the Swiss National Science Foundation

\bibliography{Fe2O3_2013bib}

\end{document}